\documentstyle[12pt,aaspp4]{article}
\newcommand{\asca}{{\sl ASCA }}
\newcommand{\ginga}{{\sl Ginga }}
\newcommand{\exosat}{{\sl EXOSAT }}
\newcommand{\rosat}{{\sl ROSAT }}

\newcommand{\egret}{{\sl EGRET }}
\newcommand{\whipple}{{\sl Whipple }}
\newcommand{\hegra}{{\sl HEGRA }}
\newcommand{\glast}{{\sl GLAST }}

\slugcomment{}

\lefthead{Kataoka et al.}
\righthead{A Study of High Energy Emission from the TeV blazar Mrk 501 
 during Multiwavelength Observations in 1996}

\begin{document}

\title{A Study of High Energy Emission from the TeV blazar Mrk 501 during 
 Multiwavelength Observations in 1996}
\author{J. Kataoka\altaffilmark{1,2}, J. R. Mattox\altaffilmark{3}, 
J. Quinn\altaffilmark{4}, H. Kubo\altaffilmark{5},  
F. Makino\altaffilmark{1}, T. Takahashi\altaffilmark{1}, 
S. Inoue\altaffilmark{6}, 
R. C. Hartman\altaffilmark{7}, 
G. M. Madejski\altaffilmark{7,8}, P. Sreekumar\altaffilmark{7,9}, 
and S. J. Wagner\altaffilmark{10}}  
\altaffiltext{1}{Institute of Space and Astronautical Science,    
3-1-1 Yoshinodai, Sagamihara, Kanagawa 229-8510, Japan}
\altaffiltext{2}{Department of Physics, University of Tokyo, 7-3-1 Hongo, 
  Bunkyo-ku, Tokyo 113-0033, Japan}
\altaffiltext{3}{Astronomy Department, Boston University, 725 Commonwealth 
 Avenue, Boston, MA 02215}
\altaffiltext{4}{Fred Lawrence Whipple Observatory, Harvard-Smithsonian Center for Astrophysics, P. O. Box 97, Amado, AZ 85645-0097}
\altaffiltext{5}{The Institute of Physical and Chemical Research, 2-1 Hirosawa, Wako, Saitama 351-0198, Japan}
\altaffiltext{6}{Department of Physics, Tokyo Metropolitan University, 1-1 
Minami-Osawa, Hachioji, Tokyo, Japan 192-0397}
\altaffiltext{7}{Laboratory for High Energy Astrophysics, NASA/GSFC, Greenbelt, MD 20771}
\altaffiltext{8}{Dept. of Astronomy, Univ. of Maryland, College Park, MD 20742}
\altaffiltext{9}{Universities Space Research Association}
\altaffiltext{10}{Landessternwarte Heidelberg, K\H{o}nigstuhl, D-69117 Heidelberg, Germany}

\begin{abstract}
We present the results of a multiwavelength campaign for Mrk 501 performed 
in March 1996 with {\sl ASCA}, \egret, {\sl Whipple}, and optical 
telescopes.  The X--ray flux observed with \asca was 5 times higher than 
the quiescent level and gradually decreased by a factor of 2 during 
the observation in March 1996.  In the X--ray band, a spectral break was 
observed around 2 keV. We report here for the first time the detection 
of high-energy $\gamma$--ray flux from  Mrk 501 with \egret with 3.5 $\sigma$ 
significance (E $>$ 100 MeV). Higher flux was also observed in April/May 
1996, with 4.0 $\sigma$ significance for E $>$ 100 MeV, and  5.2 $\sigma$ 
significance for E $>$ 500 MeV.
The  $\gamma$--ray spectrum  was measured 
to be flatter than most of the $\gamma$--ray blazars.
We find that the  multiband  spectrum in 1996 is 
consistent with that calculated from a one-zone SSC model where X--rays are 
produced via synchrotron emission, and $\gamma$-rays via inverse Compton 
scattering of synchrotron photons in a homogeneous region.  The flux 
of TeV $\gamma$--rays is consistent with the predictions of the model 
if the decrease of the Compton-scattering cross section in the 
Klein-Nishina regime is considered.  In the context of this model, we 
investigate the values of the magnetic field strength and the beaming 
factor allowed by the observational results.  We compare the March 1996 
multiwavelength spectrum with that in the flare state in April 1997. 
Between these two epochs, the TeV flux increase is  well 
correlated with that observed in keV range. 
The  keV and TeV amplitudes during the April 1997 flare are accurately
reproduced by a one-zone  SSC model, assuming that the population 
of synchrotron photons in 1996 are scattered by the newly injected 
relativistic 
electrons,  having maximum energies of $\gamma_{max}$ $\sim$ 6 $\times$ 
10$^{6}$.  However, the TeV spectrum observed during March 1996 campaign  
is flatter than predicted by our models.  We find that this cannot be 
explained by either higher order Comptonization or the contribution of 
the `seed' IR photons from the host galaxy for the first-order external 
radiation Comptonization, but we cannot exclude possible effects of 
the IR photons that may arise in the parsec-size torus postulated to exist 
in AGN.  

\end{abstract}
      

\section{Introduction}
Observations by the \egret instrument (30 MeV -- 30 GeV; Thompson et al. 1993) 
onboard the {\sl Compton Gamma-Ray Observatory} ({\sl CGRO}) reveal that  
$\gamma$--ray emission dominates the apparent power output for many blazar
type Active Galactic Nuclei (AGN).  About 50 sources detected with 
\egret in the high-energy $\gamma$--ray band are identified with blazars 
(see, e.g., Mattox et al. 1997b; Mukherjee et al. 1997).  The $\gamma$--ray 
emission extends up to TeV energies for three nearby blazars:  
Mrk 421 ($z$ = 0.031; Punch et al. 1992; Petry et al. 1996), 
Mrk 501 ($z$ = 0.034; Quinn et al. 1996; Bradbury et al. 1997; 
Aharonian et al. 1997) and 1ES 2344+514 ($z$ = 0.044; Weekes et al. 1996;  
Catanese et al. 1998). 

The overall electromagnetic spectra of blazars, when plotted in the 
$\nu$$F_{\nu}$ space, generally reveal two broad peaks; one located
between IR and X--rays, and another in the $\gamma$--ray regime 
(e.g., von Montigny et al. 1995);  in a few objects, additional components 
are observed in the IR/optical band, but those are probably not directly 
associated with the blazar phenomenon, but instead they are likely to be 
due to the isotropic emission of the 
associated AGN (e.g., Falomo et al. 1993).   
The models invoked to explain the radiative 
processes in blazars have these two components correspond to 
two different emission processes from the same population of relativistic 
particles.  The polarization observed in the radio and optical bands 
indicates that the lower energy emission is  most likely due to 
the synchrotron process, while the inverse-Compton mechanism 
is thought to be dominant for the high energy $\gamma$--ray emission 
(see, e.g., Ulrich, Maraschi \& Urry 1997).  However, the source of 
the `seed' photons for the Compton process is a matter of debate. In some 
models, it is synchrotron radiation internal to the jet --- the
Synchrotron-Self-Compton (SSC) model (Jones et al. 1974; Marscher 1980; 
K\H{o}nigl 1981; Marscher \& Gear 1985; Ghisellini \& Maraschi 1989; 
Maraschi et al. 1992; Marscher \& Travis 1996), while in others, it 
is radiation external to the jet --- the External Radiation Compton (ERC) 
model (Dermer \& Schlickeiser 1993; Sikora, Begelman, \& Rees 1994).

Blazars are highly variable from  radio to $\gamma$--ray bands:  
flares with time scales as short as 15 minutes have been observed in the 
TeV range (Gaidos et al. 1996).  The correlation of variability between 
different bands can potentially discriminate between various models.
The first TeV blazar that was observed simultaneously in multiple bands 
from radio to TeV $\gamma$--rays is Mrk 421.  The first campaign, conducted 
in 1994 (Macomb et al. 1995), revealed correlated variability between 
the keV X--ray and TeV $\gamma$--ray emission, where the $\gamma$--ray 
flux varied by an order of magnitude on a time scale of 2 days and the 
X--ray flux was observed to double in 12 hours.  On the other hand, 
the high-energy $\gamma$--ray flux observed by \egret, as well as the radio 
and UV fluxes showed less variability than the keV or TeV bands.  Another 
Mrk 421 multiwavelength campaign performed in 1995 revealed another 
coincident keV/TeV flare (Buckley et al. 1996; Takahashi et al. 
1996b).  Although the relative amplitudes of variability are
different, the UV and optical bands also showed correlation 
during the flares. These results indicate that the high energy tail of the
energy distribution of a single electron population is responsible for both 
the X--rays and TeV $\gamma$--rays (Takahashi et al. 1996a).  With the 
detection of TeV emission from Mrk 501 (Quinn et al. 1996), several multi-band 
campaigns were organized to observe this object to verify if 
such a behavior is a general property of TeV-emitting blazars, or if it is 
unique to Mrk 421.  

During the April 1997 multiwavelength campaign for Mrk 501,
both X--rays and TeV $\gamma$--rays increased by more than one order of 
magnitude from quiescent level (Catanese et al. 1997; Pian et al. 1998).
While the synchrotron emission peaked below 0.1 keV in the quiescent 
state, in 1997 it peaked at $\sim$ 100 keV --- the largest 
shift ever observed for a blazar (Pian et al. 1998);  this is in contrast 
to Mrk 421, which shows relatively little variability of the position of 
the synchrotron peak (e.g., Macomb et al. 1995).  
Mastichiadis \& Kirk (1997) suggest that the SSC scheme in a homogeneous 
region is sufficient to interpret the multiband spectrum of Mrk 421. 
They also suggest that the flare in 1994 is probably caused by a change 
in the maximum energy of electrons ($\gamma_{max}$).  The origin of the 
April 1997 flare of Mrk 501 was also investigated by Pian et al (1998).
They calculated that the variation of the $\gamma_{max}$ together with an 
increased luminosity and a flattening of the injected electron distribution 
can describe observed spectra well. Their model shows good agreement with 
the data, but does not reproduce the TeV $\gamma$-ray flux during 
the high state in April 1997.

The correlation between the high-energy $\gamma$--ray and other bands 
is important to study the radiation mechanisms of blazars.
A combined GeV and TeV spectrum can place a strong limit on 
the high energy emission and its relation with the low energy emission.  
However, observations of HBLs (High-frequency-peaked BL Lac objects) 
in the high-energy $\gamma$-ray band from 30 MeV to several GeV are difficult  
because HBLs are faint in this energy band (e.g., Mukherjee et al. 1997). 
For instance, Mrk 421 is detected at a level comparable to the \egret 
sensitivity (Fichtel et al. 1994; Thompson et al. 1995).  Catanese et 
al. (1997) present an \egret observation of Mrk 501 during the 1997
X-ray/TeV flare (Apr. 9 -- 15), and report a 1.5 $\sigma$ excess at 
the position of the source and furnish an upper limit.

In this paper we present Mrk 501 data from simultaneous March 1996 
observations by \asca(keV), \egret(GeV) and \whipple(TeV), together with 
the optical observations.  The \egret observation during this campaign 
indicates a detection with 3.5 $\sigma$ significance 
(E $>$ 100 MeV) during a 3.2 day interval of peak flux.  
Also, a month later, \egret detected Mrk 501 with 
4.0 $\sigma$ significance (E $>$ 100 MeV; 5.2 $\sigma$ for E $>$ 500 MeV).
This is the first report of the detection of 
Mrk 501 by \egret.  The observations and analysis are presented in $\S$2.
The multifrequency spectra during the campaign are described and 
compared with the April 1997 flare state and a quiescent state in $\S$3. 
We discuss the radiation mechanism of TeV blazar Mrk 501 in $\S$4, and 
present our conclusions in $\S$5.

\section{Observations and Results}
\subsection{\asca}

We observed Mrk 501 four times with \asca during 1996 March 21.3 $-$ 
April 2.9 UT. Each pointing was about 10 ksec in duration.
The observations were performed in a nominal, 1CCD mode 
for the Solid-state Imaging Spectrometer (SIS: Burke et al. 1991), 
and a nominal PH mode for the Gas Imaging Spectrometer (GIS: Ohashi et al. 
1996).  We applied standard screening procedures to the data and 
extracted the counts from a circular region centered on the target 
with a radius of 3 and 6 arcmin for SIS and GIS, respectively. 
We used standard blank sky observations to estimate the background.
The SIS light curves of these 4 observations are shown in Figure 1. 
The count rate includes the background of $\sim$ 0.01 counts s$^{-1}$.
The flux change between the observations 2 and 3 indicates variability on a  
time scale of about a day. The flux calculated from the spectrum decreased by 
a factor 2 over two weeks. No evidence was found for time variability within 
each of the four observations; the maximum $\chi^{2}$ probability of 
variability was less than  97 $\%$.

We summed all photons detected with both SISs into a single spectrum 
for each observation, because the hardness ratio (defined as the ratio of 
counting rates at 1.5 -- 7.5 keV to those at 0.7 -- 1.5  keV) did not change 
significantly.  Here we do not use the GIS data, since the efficiency of 
GISs in the low energy band (E $<$ 2 keV) is worse than that for SIS, and thus 
of less utility in a study of details of 
the spectral shape at low energies;  furthermore, for this bright source, 
and a relatively simple model, with SISs alone, we probably reached the 
level where the statistical errors are comparable to the systematic errors.  
We first fitted such combined SIS spectra (one per observation) with a single 
power law with photoelectric absorption whose column density $N_{\mbox{H}}$ 
was fixed at Galactic value, 1.73 $\times$ 10$^{20}$ cm$^{-2}$ (Elvis et al. 
1989).  However, this model did not represent any of the 4 spectra 
adequately;  a power law fitting the data above 2 keV was too steep for 
the data below 2 keV.  All the SIS data for these 4 observations showed 
such discrepancy that the reduced $\chi^2$ ranged from 1.6 to 2.0  for 
more than 300 d.o.f. The single power law model is thus rejected with higher 
than 99 \% confidence level.  

A better fit was obtained using a broken power law model with Galactic 
absorption, where the spectrum is harder below the break.  Even with this 
more complex model, we observe a clear spectral variability from one 
observation to another.  As summarized in Table 1, the spectrum steepened as 
the flux decreased from 
(9.24 $\pm$ 0.05) to (4.21 $\pm$ 0.03)$\times$10$^{-11}$ 
erg cm$^{-2}$ sec$^{-1}$ in 2 -- 10 keV.
The photon index ($\Gamma_1$) below the break energy ($E_{B}$) 
changed from 1.72 $\pm$ 0.08 to 1.89 $\pm$ 0.08 while 
the index ($\Gamma_2$) above $E_{B}$ changed from 2.15 $\pm$ 0.02 
to 2.46 $\pm$ 0.03. But the difference $\Gamma_2-\Gamma_1$ was roughly 
constant, $\sim$ 0.5, for all periods.
Although we do not use GIS data for the above fits, the parameters derived 
here are consistent when we used both GIS2 and GIS3 data. 

Figure 2 shows the correlation between the X--ray flux and the photon index. 
The maximum flux observed with \asca was five times higher than  
the quiescent level observed with \ginga (Makino et al. 1991), and 
one fifth of the highest flux ever observed (Pian et al. 1998).
As shown in Figure 2, the X--ray spectrum tends to steepen as the source 
gets fainter.  This is a general feature of HBL spectra 
(see, e.g., Giommi et al. 1990).  The spectral break around 2 keV was 
observed with both \asca and {\sl BeppoSAX} (Pian et al. 1998), while no 
spectral break was seen in the simultaneous \rosat PSPC and \ginga 
observation, when the combined 0.1 -- 20 keV spectrum was 
well-described by a power law with the photon index of 2.63 
(Makino et al. 1991).

\subsection{\egret}

We report here new \egret results for Mrk 501, for which no \egret 
detection has been reported previously.  
We have performed likelihood analysis (Mattox et al. 1996) of all viewing
periods (VP) with exposure greater than 8 $\times$ $10^{7}$ cm$^2$ s
 for $\gamma$-ray energies E $>$ 100 MeV; the results are shown in Table 2. 
Contributions from diffuse high-latitude galactic emission and
extragalactic emission (Sreekumar et al. 1998) were incorporated in the
analysis. 
Note that for VP 519.0 the result is a 4.0 $\sigma$ detection, barely 
meeting the acceptance criterion used in the construction of \egret catalogs 
(Thompson et al. 1995).

At this point, considerable caution is required, because of the
large size ( $\sim$ 5 degrees) of the \egret point spread function (PSF), 
and in this instance because of possible source confusion.  
Within 5 degrees of 
Mrk 501 are three flat spectrum radio quasars: (1) 3C 345 and (2) NRAO 512,
neither of which have ever been convincingly detected by \egret but
which are otherwise similar to the highly variable \egret-detected quasars; 
and (3) 4C +38.41, which has been detected several times by \egret, and at 
times has been very bright.  To verify that the source detected in
VP 519.0 really was Mrk 501, an analysis was done for
E $>$ 500 MeV.  Although the number of photons detected above 500 MeV 
is considerably smaller than above 100 MeV, the extent of the PSF for
the higher-energy photons is reduced by a factor of about 6 in solid
angle, enough to remove the
possibility of confusion with the other nearby objects.  The result of
the E $>$ 500 MeV analysis is a 5.2 $\sigma$ detection, and source location
contours (Figure 3) that clearly pinpoint Mrk 501 as the source of
the $\gamma$-ray emission.  The strong E $>$ 500 MeV detection suggests a very
hard spectrum, and indeed, spectral analysis gives a photon spectral
index of 1.3 $\pm$ 0.5, one of the hardest spectra ever determined with
\egret.

 The VP 516.5 \& 519.0 observations listed in Table 2
have been examined for a possible time variability.
 A Kolmogorov-Smirnov (KS) test of the cumulative number of events (with
 E $>$ 70 MeV) from the
 direction of Mrk 501 as a function of time has been performed
 (as described by Mattox et al. 1997a). For the 21 March --   3 April 1996
\egret observation (which coincides with the
\asca observation described in $\S$2.1), Kuiper's
 variant of the KS test (Press et al. 1992) indicates variability
 with 98.9 \% confidence. This interval was therefore analyzed through 
 a likelihood analysis of  four equal length intervals with
 E $>$ 100 MeV. The light curve
 thus produced is shown in Figure 1. The second sub-interval,
 24.96 -- 28.17  March 1996, showed a detection significance
 of 3.5 $\sigma$, and a flux of (32 $\pm$ 13) 
$\times10^{-8}$ cm$^{-2}$s$^{-1}$, E $>$ 100 MeV. 
The spectral photon index was measured to be 1.6 $\pm$ 0.5 in 
the 30 MeV -- 10 GeV energy range during the
 24.96 -- 28.17  March 1996 sub-interval.
A $\chi^2$ test of variability using the fluxes of these four
intervals indicates variability with 88 \% confidence. From Figure 1,
a correlation with the TeV flux appears plausible, but the \egret statistics
are too sparse for a definitive evaluation.
 For the 23 April 1996 -- 7 May 1996 exposure, Kuiper's variant of 
the KS test  indicates variability with  95 \% confidence. Higher flux 
is indicated between 23.88 and  29.13 April 1996 and 4.19 and 7.44 May 
1996.  For the  4 -- 15 April 1997 exposure,  a  likelihood analysis 
finds a 1.5 $\sigma$ indication of E $>$ 100 MeV flux. During this 
interval, Mrk 501 was observed to flare in the TeV and hard X--ray bands
(Catanese et al. 1997; Aharonian et al. 1997; Pian et al. 1998).

For VP 516.5 and VP 617.8, there are not enough photons above 500 MeV
to enable an analysis similar to that done for VP 519.0 above 500 MeV.  For
VP 516.5, the time proximity to the rather strong emission in
VP 519.0 suggests that Mrk 501 is probably the source of the 2.1 $\sigma$ 
detection, and particularly the 3.5 $\sigma$ detection in the 3.2-day
subinterval of the multiwavelength campaign, especially in view of the
TeV activity during that time.

The one-year interval between VP 519.0 and VP 617.8 prevents any similar
conclusion based on the \egret data alone; however, the strong TeV
activity during VP 617.8 again suggest that Mrk 501 is  the source of
$\gamma$--ray radiation possibly detected by \egret at that time.

\subsection{TeV observations}

Mrk 501 was observed at E $>$ 350 GeV with the \whipple {\sl Atmospheric} 
{\sl Cherenkov} {\sl Imaging}  {\sl Telescope} (Cawley \& Weekes 1995) 
from 1996 March 17 to 30 as a part of the multiband campaign.  
The data for the nights of 
March 21 and 22 were discarded due to poor weather conditions. The resulting
data set comprises 17.5 hours of good weather observations.  The data
were analyzed with the Tracking analysis using the $\gamma$--ray selection
criteria optimized on Crab Nebula data (see, e.g.,  Quinn et al. 1996). The
resulting light curve is shown in Figure 1.  The average emission
level for the observation period is 0.20 $\pm$ 0.03 photons min$^{-1}$,
approximately 15 \% that of the corresponding rate from the Crab Nebula.
This is almost double the average flux level observed from Mrk 501 in 
1995 (Quinn et al. 1996). The flux appears to be highest on the first night of
observation (0.46 $\pm$ 0.10 photons min$^{-1}$), and then decreases 
over a period
of a couple of days to a relatively constant level of 0.15 $\pm$ 0.03 
photons min$^{-1}$.  A $\chi^2$ analysis for constant emission produces a
$\chi^2$ / d.o.f of  21.5 / 11, which indicates variability at the 97 \% level
(a 2.3 $\sigma$ effect). The $\chi^2$ test was also applied to each
night's data to search for evidence of variability within a night --- 
none was found.

According to Petry et al. (1997; see also Bradbury et al. 1997),
 \hegra also observed Mrk 501 from 
March 28 to August 1996, partially overlapping our observation.
The energy threshold was 1.5 TeV. The photon index of combined spectrum 
during this 5 months was measured to be 2.5 $\pm$ 0.4 (total error) and the 
average flux was consistent with that of \whipple measured during our 
campaign in March 1996. 

\subsection{Optical observations}

A few optical observations were carried out using the 70 cm telescope
of the Landessternwarte in Heidelberg. Poor weather prohibited strictly
simultaneous observations for most of the campaign. It was possible,
nevertheless, to obtain a few measurements on March 26 and 27, 1996.
Three observations in R-band were taken in both nights, separated by
about one hour. Since no variability was found within either of the
nights within the photometric accuracy of $\sim$ 1.8 \%, the three
frames were co-added and an average flux for each of the nights was
derived.
The observed magnitude at 650 nm was $m_R$ = 11.80 $\pm$ 0.02 for March 26
and $m_R$ = 11.72 $\pm$ 0.02 for March 27. The probability of variability
in the optical band within the two nights is only 2.3 $\sigma$, and hence not
significant.
The average brightness in each night is about 30 $\%$ lower than the average
brightness in the R-band measurements from the NED data base.  Since
the absolute fluxes in any optical band depends on the correction for the
contribution of the host galaxy and a bit on the effective transmission of
standard filters and CCD response, it is not possible to judge whether this
difference is due to the source activity or systematic differences.

\section{Multiband Spectrum}
The multiband spectra of Mrk 501 taken during our campaign and at other 
times are shown in Figure 4. Open circles show the {\sl ASCA}, \egret, 
\whipple and optical data for March 25 -- 28 in 1996, when \egret detected 
the emission E $>$ 100 MeV from Mrk 501 for the first time 
at 3.5 $\sigma$ level. 
The \asca points are determined from a combined fit of data taken during 
observation  2 and  3.  The X--ray flux in the 2 -- 10 keV band was 
6.0 $\times$ 10$^{-11}$ erg cm$^{-2}$ s$^{-1}$.
The TeV count rates observed with \whipple are converted to integral fluxes 
by scaling them as a multiple of Crab Nebula count rate with assumption that 
the photon index is as same as that of Crab, 2.5 (Hillas et al. 1998). 
The TeV flux plotted in Figure 4 is 
(1.23 $\pm$ 0.6)$\times$10$^{-11}$ photons cm$^{-2}$ s$^{-1}$, 
consistent with the flux in 1995, (0.8 $\pm$ 0.1)$\times$10$^{-11}$ 
photons cm$^{-2}$ s$^{-1}$ (Quinn et al. 1996). 
The optical point is an average value of March 26 and 27, converted to the 
energy flux of (5.40 $\pm$ 0.07)$\times$10$^{-11}$ erg cm$^{-2}$ s$^{-1}$.

In order to know the change of the multifrequency spectra for various 
phases of source activity, we also plot in Figure 4 the spectra for the 
pre -- 1995 quiescent state, and the April 1997 flare.
Although the  X--ray flux observed with \asca was decreasing over a
2-week period, the long-term changes  of the X--ray spectra from 
1991 ({\sl Ginga}) to 1997 ({\sl BeppoSAX}) suggest that our March 1996 
multiwavelength campaign was conducted in the transition phase from the 
quiescent to the high state. Since a transition of the states in blazars 
is certainly not smooth or monotonous, the fact that the X--ray  
flux observed with \asca was decreasing is essentially irrelevant for 
this interpretation.

The differences in the spectral indices observed with $\asca$ 
($\Gamma_2-\Gamma_1$; see Table 1) show roughly constant value of 
$\sim$ 0.5, which is expected from the synchrotron cooling process 
(e.g., Sikora, Begelman \& Rees 1994;  Mastichiadis \& Kirk 1997).
Together with the fact that the peak of the synchrotron emission ordinarily 
exists near the UV band in the quiescent state (George et al. 1988),  
the spectral break around 2 keV observed with \asca, when the source was in 
the `intermediate' state, is considered to be the peak of the synchrotron 
emission.  The location of this break increased to $\sim$ 100 keV when the 
source was in the high state (Pian et al. 1998).  We compare the X--ray 
and TeV fluxes between three multiwavelength observations:  March 1996, 
April 7 and April 16, 1997. Compared with that in March 1996, the 2 -- 10 
keV flux was higher by a factor of 3.6 on April 7 1997, and a factor of 
8.9 on April 16 1997.  The simultaneous increase in the \whipple flux 
above 350 GeV for these two dates was by factors of 4.9 and 33 respectively.
We discuss the amplitude for the correlated variation of the keV and 
TeV band in the next section.  

\section{Discussion} 

During our campaign in March  1996, a variability of about 30 $\%$ in the 
X-ray flux was observed on time scale of a day from Mrk 501. 
Assuming that the radiation is isotropic and all photons are
created in the same region of the size implied by the variability time 
scale $t_{var}$, the optical depth to pair production for TeV and keV 
photons is estimated as $\tau$ $\gtrsim$ 30 (e.g., Mattox et al. 1997a), 
where $t_{var}$ (as measured in the observer's frame) is assumed to 
be 10$^{5}$ sec.  To avoid  absorption of $\gamma$--rays due to pair 
production, current models of blazars assume that the radiation is
Doppler boosted (e.g., Maraschi et al. 1992).  We can limit the beaming 
factor to $\delta \gtrsim$ 1.9 (assuming $H_0$ = 75 km s$^{-1}$ Mpc$^{-1}$). 
However, a stronger limit should be imposed if there were more rapid 
flux changes which we could not detect because of our sampling.
For instance, during the low state in 1986, an X-ray variability on a 
time scale of a few hours was observed from Mrk 501 by \exosat 
(Giommi et al. 1990). Quinn et al. (1998) also reported 
variability with a doubling time of 2.5 hours in the TeV band during one 
flare in June 1997.  Furthermore, there exists an evidence of correlated 
variability in U-band and TeV band, during the flare in April 1997 
(Catanese et al. 1997).
In this case, we can strengthen the limit for beaming factor to 
$\delta \gtrsim$ 6.4 for $t_{var}$ of 10$^{4}$ sec.

We apply a one-zone, homogeneous SSC model to the March 1996 multiband data, 
generally following the prescriptions of Inoue \& Takahara (1996).
A spherical geometry is adopted, with $R$ being the radius of the  
emission region. An injection of a power-law energy 
distribution of electrons 
up to a certain maximum energy 
(as expected from shock acceleration) into the radiating region
 should yield a steady-state electron distribution
with a break in its index at a characteristic energy,
determined by the balance between radiative cooling and
advective escape and/or adiabatic energy loss.
(e.g., Sikora, Begelman \& Rees 1994;  Mastichiadis \& Kirk 1997).
The resulting synchrotron emission spectrum 
should reflect this broken power-law form, with 
the electron index $p$ related to 
the spectral energy index of the emission $\alpha$ by $\alpha = (p-1)/2$.
What the observed multiband spectrum (Figure 4) actually suggests
is the presence of two breaks in the synchrotron component;
one in the IR to optical region, and the second in the X--ray band,
both with the spectral index changes of $\sim$ 0.5. 
Moreover, we were unable to adequately fit the whole spectrum 
including the $\gamma$--ray band with a model
utilizing an 
electron distribution with only one break and a maximum energy cutoff.
We therefore choose to employ 
a double broken power-law form for the electron distribution $N(\gamma)$,
in which $N(\gamma) \propto \gamma^{-s}$, $\propto \gamma^{-s-1}$ 
and $\propto \gamma^{-s-2}$ in the energy ranges
$\gamma \lesssim \gamma_{b1}$, 
$\gamma_{b1} \lesssim \gamma \lesssim \gamma_{b2}$
and $\gamma_{b2} \lesssim \gamma \lesssim \gamma_{max}$, respectively.
The injection index $s$ is taken to be $s$ = 2,
and we treat the two break Lorentz factors $\gamma_{b1}$, $\gamma_{b2}$ 
and the maximum Lorentz factor $\gamma_{max}$ as parameters.
(The particular form employed in the calculations below,
$N$($\gamma$) $\propto$ $\gamma^{-s}(1+\gamma/\gamma_{b1})^{-1} (1+\gamma/\gamma_{b2})^{-1}$,
was chosen simply for convenience;  using sharp breaks instead 
produces only small changes in the resulting spectrum.)
Such a second break (or equivalently, a broad cutoff) in the distribution
may effectively result from incorporating a more realistic geometry
and/or detailed kinetic evolution of the particle distribution
in the downstream region of the shock front
(see, e.g., the discussion in Kirk, Rieger \& Mastichiadis 1998).
However, a deeper inquiry into this aspect is beyond the scope of this paper;
here it is only introduced in order to provide a better description of the 
data.

Synchrotron emission including self-absorption  
is calculated using the standard spherical solution to the radiative 
transfer equation,
\begin{equation}
L_{sync} (\nu) = 4 \pi^2 R^2 \frac{j_\nu}{\alpha_\nu} 
(1 - \frac{2}{{\tau_{\nu}}^2}[1 - e^{ - \tau_{\nu}}(\tau_{\nu} + 1 ) ] ) 
\end{equation}
where $j_{\nu}$ and $\alpha_{\nu}$ 
are respectively the emission and absorption coefficients for synchrotron 
radiation (e.g., Blumenthal \& Gould 1970; Gould 1979;   
Bloom \& Marscher 1996). 
$\tau_{\nu}$ is the optical depth in the blob along the line of sight  
and expressed as $\tau_{\nu}$ = 2$\alpha_{\nu}$$R$. 
The electron pitch angle with respect to the magnetic field 
is set to $\theta$ = $\pi$/2. 

We calculate the inverse Compton emission
incorporating the effects of cross section reduction in the Klein-Nishina 
regime. The differential photon production rate $q(\epsilon)$ is
\begin{equation}
q(\epsilon) = \int d \epsilon_0 n(\epsilon_0) \int d \gamma N(\gamma) 
C(\epsilon, \gamma, \epsilon_0)
\end{equation}
where $\epsilon_0$ and $\epsilon$ are respectively 
the soft photon energy and the scattered photon energy in units of $m_e c^2$,
$n(\epsilon_0)$ is the number density of soft photons per energy interval,
and $C(\epsilon,\gamma,\epsilon_0)$ is the Compton kernel of Jones (1968).
To be exact, $n(\epsilon_0)$ varies depending on the position in the emission 
blob and we have to take this effect into account. 
Approximately, we calculated $n(\epsilon_0)$ 
at the center of the blob with the correction factor of $C_{corr}$ = 0.75 
(Gould 1979). Thus we have 
\begin{equation}
n(\epsilon_0) = \frac{4 \pi}{h c \epsilon_0} C_{corr} \frac{j_\nu}{\alpha_\nu}
 (1 - e^{-\alpha_\nu R})
\end{equation}
where $h$ is the Planck constant.
The convolution in equation (2) is performed under the condition 
\begin{equation}
\epsilon_0 \le \epsilon \le \gamma\frac{4\epsilon_0\gamma}{1+4\epsilon_0\gamma}
\end{equation}
from the kinematics of electron-photon scattering.   

We applied this model to the observed spectrum of Mrk 501. 
To approximate the values for magnetic field and beaming factor, 
we first place constraints on the allowed parameter space similar to 
Bednarek \& Protheroe (1997), calculated  for the case of Mrk 421.
In the following prescriptions, we basically assume the 
parameter space for the March 1996 data. However, because of the lack of  
the information in the hard X-ray band and the high energy tail of the 
TeV spectrum during this campaign, we also use the non-contemporaneous data 
sets for the calculation. 
  
Since the \ginga data in the quiescent state shows no spectral cut off  
below 20 keV (Figure 4), it is probable that the synchrotron spectrum in 
March 1996 also extends above 20 keV.
We assume here the maximum synchrotron energy at 
50 keV. We thus  obtain 
$\gamma_{max}$ = 3.2$\times$10$^{6}$ $B^{-0.5}$ $\delta^{-0.5}$.
The Compton-scattered spectrum extending to higher than 10 TeV 
(Aharonian et al. 1997) gives us the limitation of  
 $\gamma_{max}$ $m_ec^2$ $\ge$ 10 TeV/$\delta$. 
Using both equations, we can limit the magnetic field  to 
 $B$ $\le$ 2.6$\times$10$^{-2}$ $\delta$ Gauss.
This limit is represented by line (I) in Figure 5. 

The second  constraint is derived using the ratio of the synchrotron 
and Compton luminosities.  
Here we can use a relation for the Thomson regime since the Compton luminosity 
is well approximated as that for the  \egret  range and the high-energy 
$\gamma$-ray spectrum shows no sign of Klein-Nishina cutoff in the   
MeV -- GeV band (Figure 4).
Thus the energy density of the synchrotron radiation $u_{sync}$ is related 
to the magnetic field density $u_{B}$ as 
$u_{B}$ = $u_{sync}$ ($L_{sync}$/$L_{SSC}$), 
where $L_{sync}$ and $L_{SSC}$ 
is the observed synchrotron/Compton luminosity. 
Since $u_{sync}$ is represented as 
\begin{equation}
u_{sync} = \frac{L_{sync}}{4 \pi R^2 c \delta^{4}}
\end{equation} 
we have the relation  
\begin{equation}
u_B = \frac{{d_L}^2}{R^2 c \delta^4} \frac{{l_{sync}}^2}{l_{SSC}}
\end{equation} 
where $d_L$ is the luminosity distance, $l_{sync}$ and $l_{SSC}$ are 
energy flux in observer's frame. From Figure 4, we can approximate 
$l_{sync}$ and $l_{SSC}$ $\sim$ 10$^{-10}$ erg cm$^{-2}$ s$^{-1}$.
The upper limit for $R$ is also  given as 
\begin{equation}
R \le c t_{var} \delta
\end{equation} 
where $t_{var}$ is the minimum variability time scale in 
the observer's frame.  
Using equation (6) and (7), with $t_{var}$ expressed in seconds, we obtain 
 $B$ $\ge$ 4$\times$$10^6$ $t_{var}^{-1}$ $\delta^{-3}$ Gauss.
This limit is represented by line (II) in Figure 5. 

The third constraint is derived from the evaluation of a variability 
time scale and synchrotron cooling time scale ($t_{sync}$). 
Although the minimum variability time scale must be determined by the 
balance between radiative cooling and adiabatic energy loss or advective 
escape (e.g., Sikora, Begelman \& Rees 1994;  Mastichiadis \& Kirk 1997), 
the synchrotron emission can be one of the most dominant cooling processes; 
the X-ray spectrum of Mrk 501 which tends to steepen as the source gets 
fainter supports this assumption (see Figure 2).   
For the electrons of large $\gamma$ ($\gamma$ $\ge$ $\gamma_{b1}$), 
$t_{sync}$ is smaller than the source crossing time scale $R/c$.  
Combining this with equation (7), we obtain 
$t_{var}$$\delta$ $\ge$ $t_{sync}$. 
Since $t_{sync}$ is represented as 
\begin{equation}
t_{sync} = \frac{3 m_ec}{4 u_B \sigma_T \gamma }
\end{equation} 
where $\sigma_T$ is the  Thomson cross section,
   it becomes minimum at $\gamma$ = $\gamma_{max}$. The lower 
limit for $B$ is thus calculated as $B$ $\ge$ 3.9$\times$10 $t_{var}^{-2/3}$ 
$\delta^{-1/3}$ Gauss. 
This limit is represented by line (III) in Figure 5. 

The final constraint is that from the evaluation of the variability 
time scale and Compton cooling time scale ($t_{SSC}$). 
Even for the creation of TeV photons, scatterings in the Thomson regime can be 
dominant if $\gamma_{max}$$h$$\nu$ $<$ $m_e$$c^2$$\delta$ and 
$\gamma_{max}^2$$h$$\nu$ $>$ 1 TeV are satisfied at the same time, 
where $h$$\nu$ is the soft photon energy in the observer's frame.
These inequalities have a self-consistent solution for $h$$\nu$ $<$ 
$\delta^{2}$/4 eV, under the condition of $B$ $<$ 5$\delta$ Gauss.
We then approximate $t_{SSC}$ using equation (8), replacing the magnetic 
field density by the synchrotron photon density. 
To estimate the value of  $u_{sync}$, $L_{sync}$ $\sim$ 10$^{45}$ 
erg s$^{-1}$ is used here. Thus $t_{SSC}$ at $\gamma$ = $\gamma_{max}$ is 
calculated as 3$\times$10$^{-12}$ $t_{var}^2$ $B^{0.5}$$\delta^{6.5}$, 
where we assume $R$ as $c$$t_{var}$$\delta$,  rather than using 
inequality (7). 
This is not true in the strict sense, although it is plausible if 
$t_{SSC}$ at $\gamma$ = $\gamma_{max}$ is much shorter than $t_{var}$
$\delta$. The previous discussion for $t_{var}$$\delta$ $\ge$ $t_{sync}$ 
 supports our assumption if $u_B$ $\sim$ $u_{sync}$.
The magnetic field is then calculated to be 
$B$ $\le$ 9$\times$10$^{22}$ $t_{var}^{-2}$ 
$\delta^{-11}$ Gauss.  This limit is represented by line (IV) in Figure 5.

In Figure 5, we show the allowed values of $B$ as a function of $\delta$ 
for the case of Mrk 501 data sets. These four limits are derived for 
$t_{var}$ of 10$^5$ sec, which is observed with \asca in March 1996. 
We also calculate the limits that corresponds to 
the minimum time scale $t_{var}$ of 10$^4$ sec, for more rapid flux changes 
that we could not detect because of our sampling.
The input parameters for our model are determined at the center of the 
overlapped region for those different time scales 
: $B$ = 0.2 Gauss, $\delta$ = 15 and $R$ = 4.5 $\times$ $10^{15}$ cm. 
The electron distribution is then fixed as: $\gamma_{min}$ = 1.0, 
$\gamma_{b1}$ = 1.2 $\times$ $10^{4}$, 
$\gamma_{b2}$ = 3.7 $\times$ $10^{5}$  and 
$\gamma_{max}$ = 1.8 $\times$ $10^{6}$. 
As shown in Figure 4, the whole spectrum is adequately represented, except 
for the IR/optical feature near 10$^{14}$ Hz and a discrepancy in the TeV 
spectrum. One can  also see the discrepancy in the radio band, but this can 
be an effect of at least a part of the emission arising at a larger 
distance than the location of the keV/TeV emitting region.  
The one-zone models cannot account for the low-energy emission, which is 
thought to be produced in a much larger region of the source (e.g., 
Marscher et al. 1980; see also Pian et al. 1998). However, investigation
of this low-energy discrepancy is beyond the scope of this paper; in the 
following, we concentrate on the discrepancy in the higher energy band 
where the one-zone models are thought to be a reasonable  approximation 
to account for the power output of blazars.

The predicted photon index in the \hegra band is 3.8, while the observed 
value is  2.5 $\pm$ 0.4 (Petry et al. 1997; Bradbury et al. 1997). 
The photon index in the \egret energy range (30 MeV -- 10 GeV) is predicted 
to be 1.8  which  is consistent with the observed value of 1.6 $\pm$ 0.5.  
The rest--frame luminosity is also calculated to be 
$L'_{sync}$ = 2.2 $\times$ $10^{40}$ erg s$^{-1}$ and 
$L'_{SSC}$ = 1.8 $\times$ $10^{40}$ erg s$^{-1}$. From integration of 
$\gamma$$N(\gamma)$, the energy density of relativistic electrons in the 
source frame is estimated to be 0.16 erg cm$^{-3}$ and Compton optical 
depth is $\tau_c$ = 6.3 $\times$ $10^{-5}$.  

It is worthwhile to compare the overall spectrum, and the parameters 
derived by us to those obtained for Mrk 421 (see, e.g., Takahashi 
et al. 1996a).  Observationally, the spectrum of Mrk 421 also shows a  
spectral break around  1 keV, but its energy does {\sl not} seem to vary more 
than a factor of a few when the source enters a flare state, while Mrk 501 
shows a change of a factor of $\sim 100$.  Our modelling indicates that the 
derived magnetic field $B$ is comparable between the two sources, on the 
order of 0.2 Gauss, but the main difference appears to be the energy of 
electrons radiating at the peak of the synchrotron component during the 
flare, $\gamma_{max}$, which for Mrk 421 is a few $\times$ $10^5$ to $10^6$, 
while for the flare of Mrk 501, it is larger, several $\times$ $10^6$ or 
greater (e.g., Inoue \& Takahara 1996; Pian et al. 1998; see also below).  
While for Mrk 421, an independent verification of these parameters exists 
via the comparison of the Klein-Nishina limit to the implications of the 
spectral variability observed in X--rays (e.g., Mastichiadis \& Kirk  1997; 
Bednarek \& Protheroe 1997), 
we do not have a well-sampled 
X--ray light curve for Mrk 501, and thus cannot make such a comparison here.  

In any case, those large differences indicate that the overall 
electromagnetic spectra of HBLs are qualitatively similar, but 
peaking at different energies, which is likely to be a result of 
different energy distribution of radiating electrons.  
Kubo et al. (1998) found that $B$ is similar over a large 
range of blazar classes, and the difference of the 
overall spectra -- where the blazars associated with quasars (showing 
emission lines) have synchrotron components peaking in the infrared, as 
opposed to HBLs, where these peaks are in the X--ray regime -- 
can be also explained by  lower $\gamma_{max}$ in quasar-type blazars 
than in HBLs.  However, it is not possible 
as yet to determine if this is intrinsic, i.e. related to different 
conditions for the particle acceleration process, or if some external 
conditions (such as the strong diffuse 
radiation field) play an important role in the particle population inferred 
from the observations.  

We next investigate whether the one-zone SSC model might be helpful to 
understand the large flare observed in April 1997.  Inspection of Figure 4 
suggests that two different components probably contribute to the synchrotron 
spectrum, one a steady component, and another, variable,  
responsible for the sudden increase of the highly energetic electrons, 
when the source goes into the flare state.  We set $\gamma_{max}$ to 
5.5 $\times$ $10^{6}$ for the April 7 data and 7.0 $\times$ $10^{6}$ for 
the April 16 data.  The parameters such as region size, magnetic field, 
and beaming factor are the same as that for March 1996.  The normalization 
was determined from the synchrotron spectra obtained during the 
{\sl BeppoSAX} observations (Pian et al. 1998).  
For these flare data, a single break 
population of electrons is adequate to account for the multiband spectrum 
($\gamma_{b1}$ $\sim$ $\gamma_{b2}$).  

We first assumed that the newly 
injected electrons scatter only the photons emitted by the electron  
population associated with the flare.  However, the calculated  
flare amplitude of the TeV flux was substantially lower than that observed.  
We next 
assumed that there exists a steady  population of synchrotron photons, 
with the same level as that in 1996,  and those are scattered by the 
flare electron population.  In this case, the flare amplitude of the 
TeV flux is quite well reproduced as shown in Figure 4.  The photon 
index in the \hegra region is predicted to be 2.6  for April 7 1997,  
comparable with the observational data of 2.49 $\pm$ 0.11 $\pm$ 
0.25 (Aharonian et al. 1997).  This result is consistent with Pian et 
al. (1998), where the scattering of the  steady component by the 
flare electron population is necessary to account for the flare 
amplitude of the TeV flux.  This may imply that the steady and 
the flare components arise from close parts of the jet, having some 
interaction with each other.  Unfortunately, the observed data during 
the flare are too sparse to discriminate whether the two emission 
components originate in the same region or different regions of the jet.
More precise observations in various energy bands (and, in particular, 
better sampling) are necessary to further understand this problem.

The relatively flat shape of the spectrum observed with \hegra in 
March 1996 is discrepant 
with our one-zone model.  We investigated two possible causes of this 
discrepancy.  The first is the external soft photons from the host 
galaxy or obscuring torus and the other is the multiple Comptonization.  
The integration of the stellar light from the galaxy can  
be the origin of the IR/optical feature present around 10$^{14}$ Hz 
(see Figure 4; Falomo et al. 1993). These photons are distributed 
isotropically in the observer's frame and are injected externally into 
the emission region. The IR photons are important since 
they can be scattered to TeV energies 
($\gamma_{max}$ $\sim$ 10$^6$) without entering the Klein-Nishina regime. 
Additionally, Comptonization of such external photons  can be more effective 
than SSC, because of the anisotropy of injected photons in the frame 
of the jet (Sikora et al. 1994;  Dermer 1995; Dermer et al. 1997). 
To account for the flat TeV spectrum with such external radiation, 
the observed external Compton luminosity ($L_{ext}$) must be at 
least comparable with the synchrotron luminosity ($L_{sync}$).  
This means that the energy densities of magnetic field  and 
external photons have the relation 
$u_{ext}$ $\sim$ $u'_{ext}$/$\delta^2$ $\gtrsim$ $u_{B}$/$\delta^4$, where 
$u_{ext}$  is the density of external photons in the observer's frame 
and $u'_{ext}$ in the jet frame.   For our model of $B$ = 0.2 Gauss and 
$\delta$ = 15, we obtain the constraint of $u_{ext}$  $\gtrsim$ 10$^{-8}$ erg 
cm$^{-3}$.  The typical values for the total 
luminosity of the galaxy $L_{gal}$ = 10$^{44}$  erg s$^{-1}$  and 
effective radius $R_{eff}$ = 1 kpc (e.g., Binney  \& Tremaine 1987) 
give the energy density of $u_{ext}$ $\sim$ 3 $\times$ 10$^{-11}$ erg 
cm$^{-3}$ ---  much smaller than the lower limit given above and thus the 
galactic IR photons aren't expected to affect the TeV spectrum.  

However, if the IR photons arise in a region comparable to a parsec-scale 
torus, postulated to exist in Seyfert galaxies (e.g., Antonucci \& Miller 
1985;  Krolik \& Begelman 1986) --- then, such a photon field may be 
important (e.g., Sikora et al. 1994).  These photons can be emitted from 
the dust around the nucleus, but may be hidden by the strong jet emission 
in the observer's frame;  in fact, our spectrum (see Figure 4) does not 
exclude such emission. We can easily  predict that if the total luminosity 
of such IR photons exceeds $L_{dust}$ $\gtrsim$ 10$^{43}$ erg s$^{-1}$,
 it can contribute sufficient seed photons that subsequently would be 
Comptonized to dominate the TeV spectrum.
The precise (sub-arcsec) IR/optical imaging of Mrk 501 
should limit the spatial extent of the `IR--bump' mentioned above.  

The second possible cause of the TeV spectral discrepancy is
the contribution of photons produced by multiple Compton scattering 
(e.g., Band et al. 1985, 1986;  Bloom \& Marscher 1996).  
In considering the second-order scattering, the same formula (2) 
is used, except that the incident photons are now the first-order inverse 
Compton photon density (Bloom \& Marscher 1996). 
Our calculation shows the contribution from multiple Compton scattering 
is  greatly suppressed due to the Klein-Nishina effect. The flux ratio 
for Comptonization of first and second order is $f_{2nd}$/$f_{1st}$ 
$\lesssim$ 10$^{-3}$ at 1 TeV.  Thus, the second order SSC process 
does not make a significant contribution to the spectrum, and we are 
unable to explain the relatively  flat TeV spectrum in March 1996.

A beaming factor much larger than 10 could also make TeV spectrum flatter.  
However, our previous calculation shows that at $\delta$ $\gtrsim$ 40, 
Compton cooling time for highly energetic electrons becomes larger than 
the source crossing time scale $R/c$ and thus 
no clear break should appear in the electron population. 
Also, a beaming factor much larger than 10 (e.g., $\delta$ $\gtrsim$ 100) 
doesn't reproduce the GeV/TeV flux ratio  of $f_{EGRET}$/$f_{Whipple}$ 
$\sim$ 10, but makes both fluxes comparable because most of the scattering
 will take  place in the Thomson regime.

Although the boundary for the allowed parameter space could  be changed 
if variability on shorter time scales is found (see Figure 5), a more 
tightly constrained TeV spectral index (through better statistics and less 
systematic error) will be of great value to make further progress on 
modelling of this source.  
Together with 
more refined SSC models such as Marscher (1980) and Ghisellini \& Maraschi 
(1989), we have shown that an exact evaluation  of Comptonized emission 
from external IR photons, such as dust emission 
around the nuclei and accretion disk photons, 
may be useful for the understanding of the origin of the flat TeV spectrum.  
Our model here is constructed with  no time dependence. 
To be exact, this model should be expanded to include 
the full time dependent evolution of the electron and photon spectra
(e.g., Mastichiadis \& Kirk  1997). 
The time dependent treatment for Mrk 501 flare will be discussed elsewhere. 
   
\section{Conclusion}
We have conducted a multiwavelength campaign for the TeV  blazar Mrk 501 
in March 1996.  In the \egret data, we found 3.5 $\sigma$ detection 
(E $>$ 100 MeV) of high-energy $\gamma$-rays.  
Higher flux was also observed in April/May
1996,  with 4.0 $\sigma$ significance for E $>$ 100 MeV, and  5.2 $\sigma$
significance for E $>$ 500 MeV. The maximum flux 
observed with \asca was 5 times higher than the quiescent level 
and gradually decreased by a factor of 2 as the spectrum steepened. 
The spectra obtained with \asca show a `break' around 2 keV, which 
is interpreted as  the peak of the synchrotron emission.  
Comparing our results with the observations in April 1997, there were 
notable changes in overall spectra, such as a 2 decade shift of 
the peak frequency of the synchrotron component, a change of 
peak flux in the keV region by  a factor of 9, and a flux change in the 
TeV region by a factor of 33. To investigate a correlated variation of 
spectra in the keV and TeV region, we applied a one-zone, homogeneous 
SSC model with an electron distribution having two break points. 
The allowed parameter space ($B$, $\delta$) was systematically 
investigated to obtain values for the model parameters.
The spectrum in March 1996 was consistent with parameters 
$\gamma_{max}$ = 1.8 $\times$ 10$^6$, $B$ = 0.2 Gauss, $\delta$ = 15, and 
$R$ = 4.5 $\times $10$^{15}$ cm. The correlation of the keV/TeV flare 
amplitudes in April 1997 was also reproduced with the  assumption that an  
additional electron population, with higher maximum energies 
($\gamma_{max}$ $\sim$ 6 $\times$ 10$^6$) was newly injected and scattered 
the photons emitted from the steady component, assumed to be the same 
level as that in 1996.  We found that 
the scattering of the synchrotron photons emitted from only the flare 
electron population is inadequate in our model.  
The flat TeV spectrum observed with \hegra in March 1996 is discrepant
with our model. We considered the contribution from the galactic 
IR photons, and higher order Comptonization, and found that neither
could make the TeV spectrum flatter.  
A possible contribution from a parsec-scale torus, postulated to exist in  
Seyfert galaxies, was also investigated.  
We need better quality data, especially in the hard X--ray 
and $\gamma$--ray regimes, to better understand 
the emission mechanism. Observations with a next generation of satellites, 
featuring improved sensitivity at hard X--ray/$\gamma$--ray energies, such 
as {\sl Astro -- E} and \glast, are expected to bring valuable information. 

\acknowledgments
This research has made use of the NASA/IPAC Extragalactic Database (NED) which 
is operated by the JET Propulsion Laboratory, Caltech, under contact with the 
national Aeronautics and Space Administration, and made use of data from the 
University of Michigan Radio Astronomy Observatory which is supported by 
the National Science Foundation and by funds from the University of Michigan.
We would like to thank Dr. A. P. Marscher, Dr. T. C. Weekes and an anonymous 
referee for their constructive comments and discussions.  
We also thank Dr. M. Tashiro and 
Dr. T. Mihara for communication of the \ginga data in 1991. 
J. R. Mattox acknowledges support from NASA grant NAC 5-3384.

\clearpage

\begin{deluxetable}{crrrrrrrrrr}
\footnotesize
\tablecaption{Fit result of \asca spectrum\label{tbl-1}}
\tablewidth{0pt}
\tablehead{
\colhead{Obs.ID}
& \colhead{time(UT in 1996)}
& \colhead{$\Gamma_1$\tablenotemark{a}} &  \colhead{$E_B$\tablenotemark{a}}
& \colhead{$\Gamma_2$\tablenotemark{a}} &  \colhead{Flux\tablenotemark{b}}
& \colhead{$\Gamma_2$$-$$\Gamma_1$} &  \colhead{ $\chi^{2}$ / (d.o.f)} }

\startdata
1 &Mar 21.26--21.50 & 1.72$\pm$0.02  &1.72$\pm$0.08 & 2.15$\pm$0.02 &9.24$\pm$0.05 & 0.43$\pm$0.03 & 1.00(379)
\nl
2 &Mar 26.12--26.42 &1.77$\pm$0.03 & 1.67$\pm$0.08 &2.22$\pm$0.02  &7.03$\pm$0.04  & 0.45$\pm$0.04  &0.92(360)
\nl
3 &Mar 27.70--27.95  &1.89$\pm$0.03 &1.67$\pm$0.07 &2.35$\pm0.02$ &5.20$\pm$0.03 &0.46$\pm$0.04 &  1.04(329)
\nl
4 &Apr 2.69--2.93 &1.89$\pm$0.03 &1.65$\pm$0.07 & 2.46$\pm0.03$ &4.21$\pm$0.03 & 0.57$\pm$0.04 &1.04(313)
\nl
\enddata

\tablenotetext{a}{
Fit model is broken power law with Galactic absorption.
$\Gamma_1$ and $\Gamma_2$ are photon indices
in lower and higher energy band, respectively.
$E_{B}$ is break energy in keV. SIS0 and SIS1 data are combined.}

\tablenotetext{b}{ 2--10 keV flux in units of 10$^{-11}$erg cm$^{-2}$ 
s$^{-1}$.}

\tablenotetext{}{All errors are 1 $\sigma$. }
\end{deluxetable}

\begin{deluxetable}{clcccrrrrrr}
\footnotesize
\tablecaption{\egret results for Mrk 501\label{tbl-2}}
\tablewidth{0pt}
\tablehead{
\colhead{\egret VP\tablenotemark{a}}& \colhead{Interval} &
\colhead{Aspect\tablenotemark{b}}
& \colhead{Exposure\tablenotemark{c}} & \colhead{Significance
($\sigma$)} 
& \colhead{Flux\tablenotemark{d}}} 

\startdata
 9.5 &  12-Sep-1991 ---  19-Sep-1991 &  3.3$^\circ$ &  2.28  &0.5 &
$<$11\nl
 201.0 &  17-Nov-1992 ---  24-Nov-1992 &  2.5$^\circ$ &  1.12  &1.7 &
9$\pm$6\nl
 202.0 &  24-Nov-1992 ---  01-Dec-1992 &  5.8$^\circ$ &  1.06  &1.1 &
8$\pm$7\nl
 516.5 &  21-Mar-1996 ---  03-Apr-1996 &  3.1$^\circ$ &  1.47  &2.1 &
10$\pm$5\nl
       & (25-Mar-1996 ---  28-Mar-1996)&              &  0.46  &3.5 &
32$\pm$13\nl
 519.0 &  23-Apr-1996 ---  07-May-1996 &  1.2$^\circ$ &  2.10  &4.0 &
18$\pm$5\nl
       &       (E$>$500 MeV)                     &              &  2.97
&5.2 & $^e$6$\pm$2\nl
 617.8 &  04-Apr-1997 ---  15-Apr-1997 &  3.0$^\circ$ &  0.82  &1.5 &
9$\pm$7\nl
\enddata

\tablenotetext{a}{\egret Viewing Period.}
\tablenotetext{b}{Angle between the source and the \egret instrument
axis.}
\tablenotetext{c}{exposure in unit of $10^{8}$ cm$^2$ s.}
\tablenotetext{d}{Flux (E$>$100 MeV) in units of $10^{-8}$
cm$^{-2}$s$^{-1}$. All errors are 1 $\sigma$.}
\tablenotetext{e}{Flux (E$>$500 MeV) in units of $10^{-8}$
cm$^{-2}$s$^{-1}$. The error is 1 $\sigma$.}
\end{deluxetable}

\begin{figure}
\epsscale{0.8}
\plotone{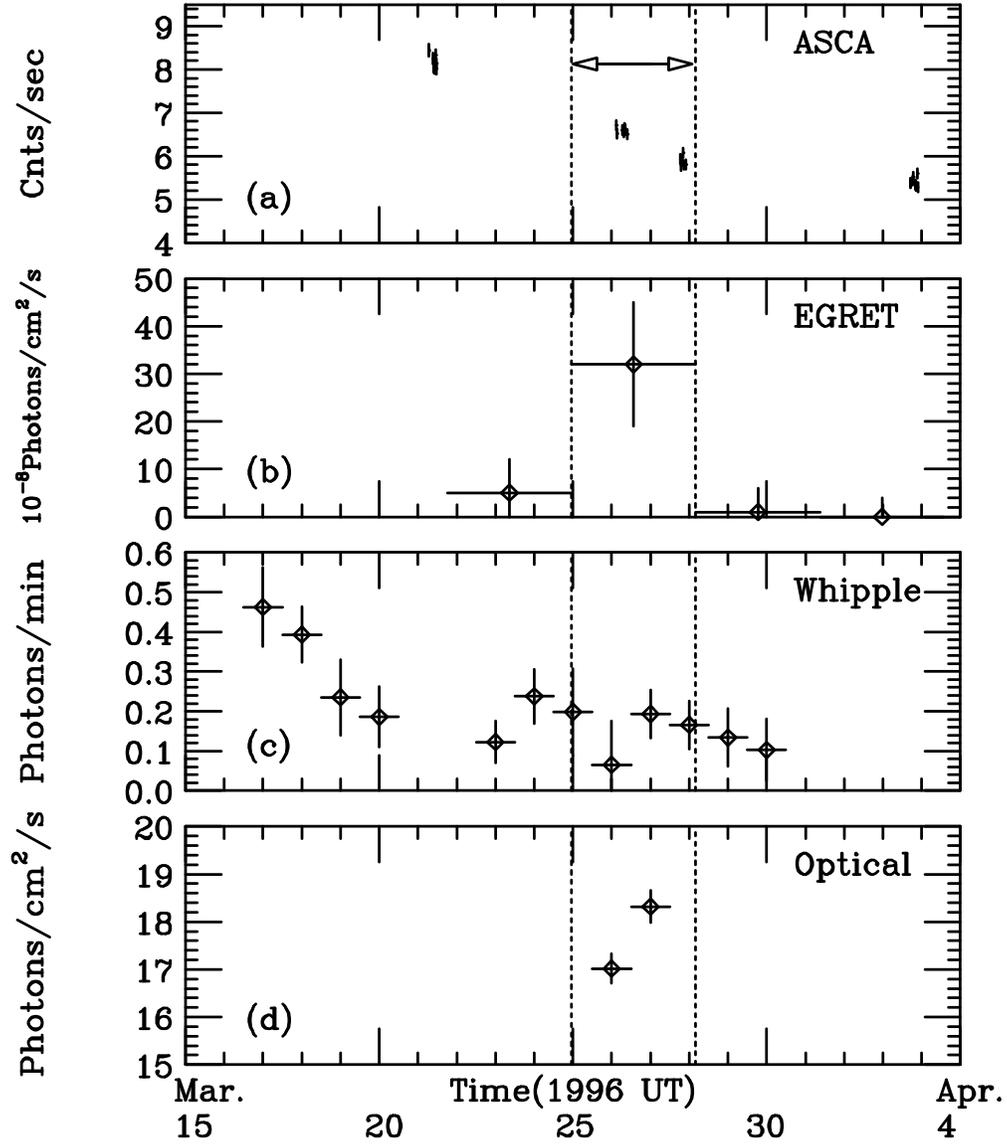}
\figcaption{Time history of Mrk 501 during the March 1996 campaign.
(a): X--ray ({\sl ASCA}: 0.7 -- 10 keV), 
(b): GeV ({\sl EGRET}: 100 MeV -- 10 GeV), 
(c): TeV ({\sl Whipple}: above 350 GeV) and (d): Optical (R -- band: 650 nm).
The \asca count rates are from the summed SIS0 and SIS1 data, 
extracted from a circular region centered on the target with a radius of 3 
arcmin. 
The time interval marked with arrows is  when \egret detected Mrk 501 
at 3.5 $\sigma$ significance for the first time. All errors 
are 1 $\sigma$.\label{fig.1}}
\end{figure}

\clearpage

\begin{figure}
\epsscale{0.8}
\plotone{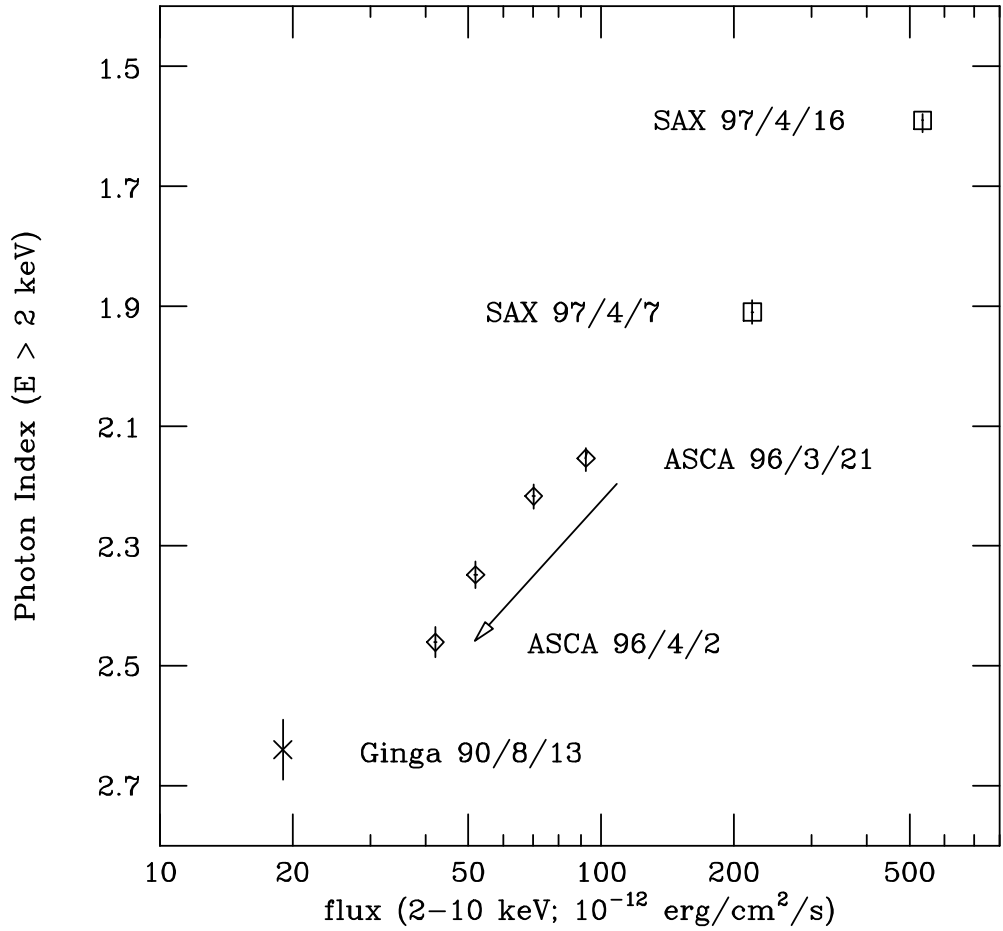}
\figcaption{Correlation between 2--10 keV flux and spectral photon  
index above 2 keV for Mrk 501.
\ginga and {\sl BeppoSAX} results are respectively from Makino et al. (1991) 
and Pian et al. (1998). All errors are 1 $\sigma$. 
Arrow indicates an evolution during the \asca observation 
in March \& April 1996.\label{fig.2}}
\end{figure}

\clearpage

\begin{figure}[htp]
         \includegraphics{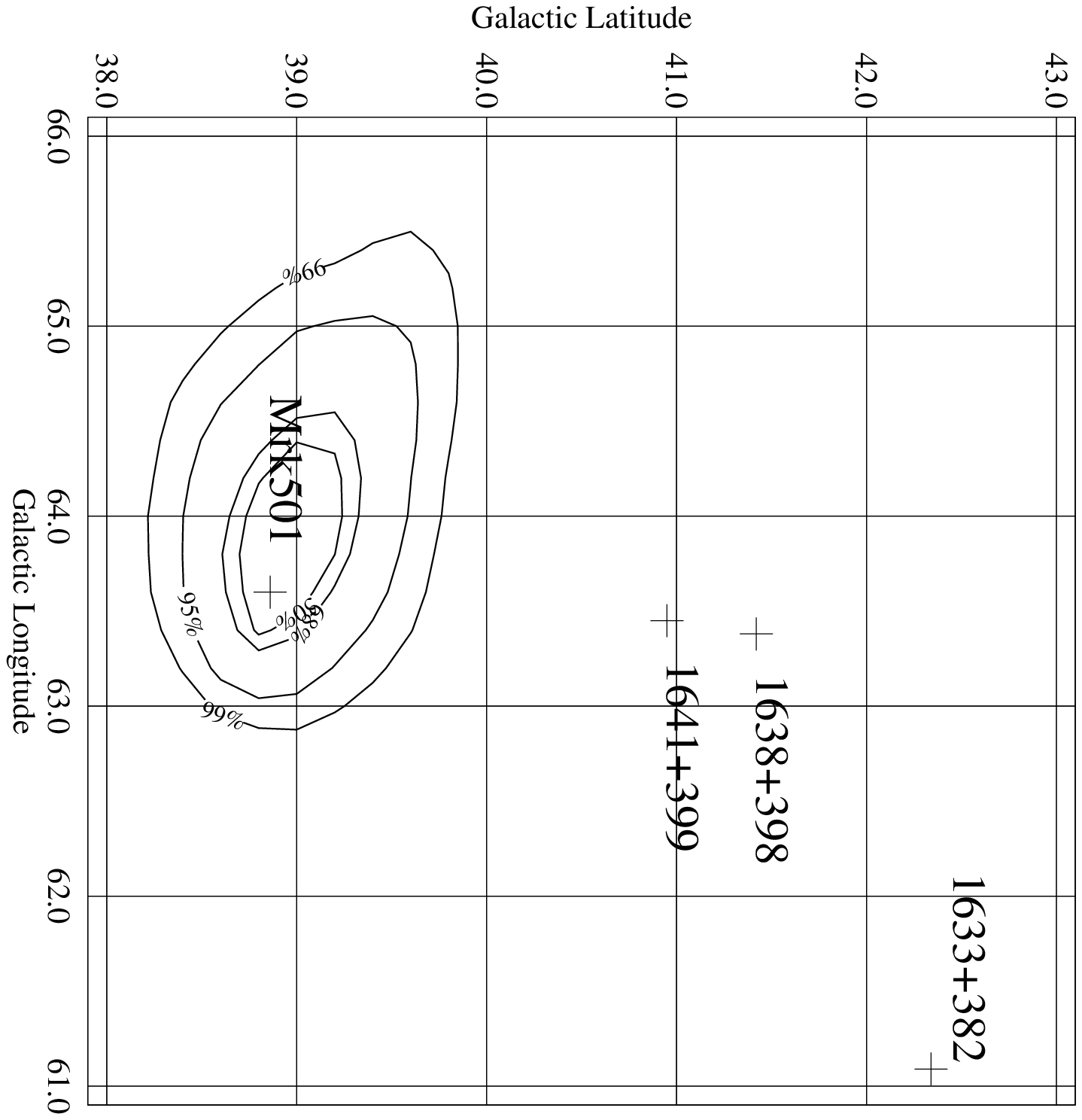} 
        \vspace*{4.0cm}  

\vspace{10.8cm}      
\figcaption{Maximum likelihood map of \egret observations using 
photons above 500 MeV. The smaller PSF at higher energies yields good 
positional identification with Mrk 501 while clearly ruling out association 
with other nearby blazars.\label{fig.3}}
\end{figure}

\clearpage

\begin{figure}
\epsscale{1.0}
\plotone{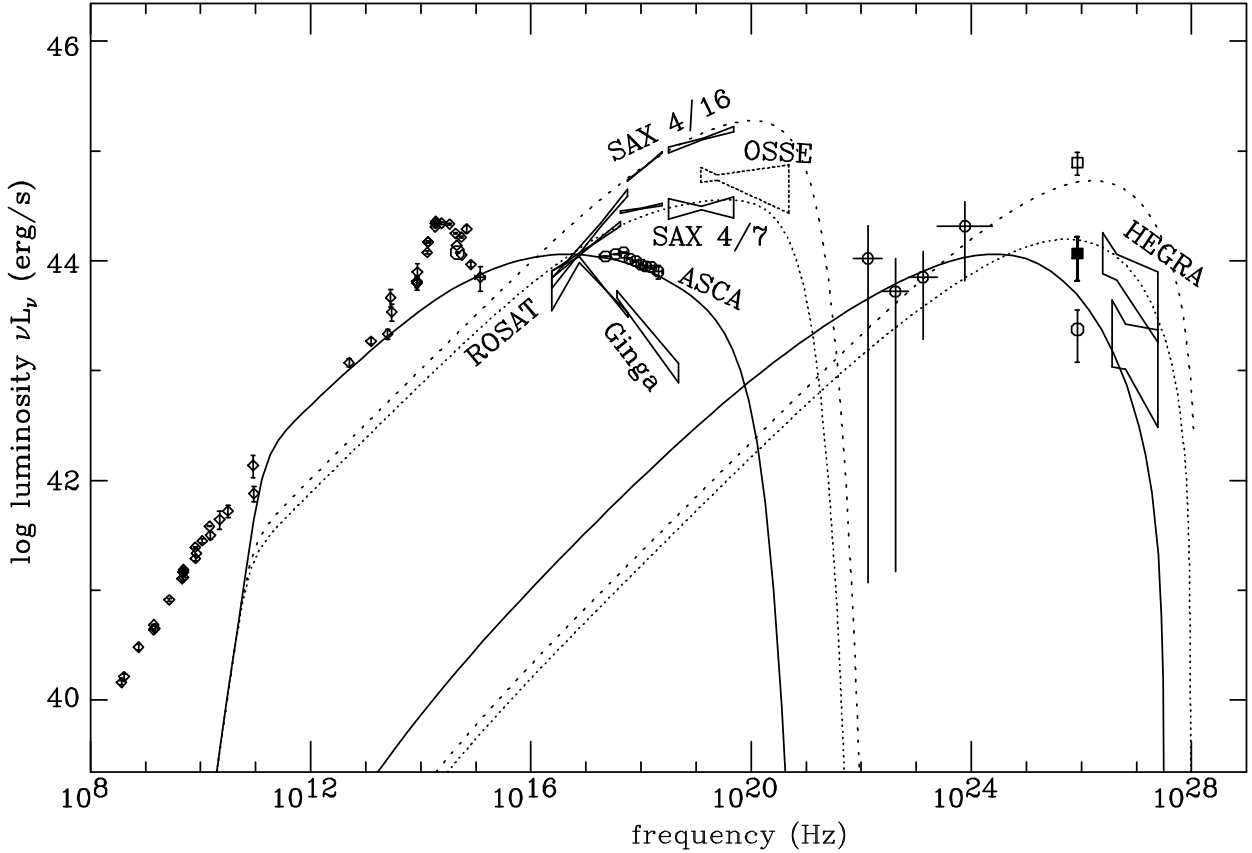}
\figcaption{Multiband spectrum of Mrk 501. 
Circle: March 25 -- 28 1996 campaign. Open Square: \whipple flux  on April 16, 
1997. Filled Square: \whipple flux on April 7,  1997.
Diamond: non-simultaneous data from the NED data base.
\rosat (Comastri et al.1997), \ginga (Makino et al. 1991), 
{\sl BeppoSAX} (Pian et al. 1998), {\sl OSSE} (Catanese et al. 1997), 
\egret (this work), 
\whipple (Quinn et al. 1996; Catanese et al. 1997) and 
\hegra (Bradbury et al. 1997; Aharonian et al. 1997).
{\sl OSSE}  data are average flux from April 9 to April 15, 1997. 
The lower \hegra plots are  averages  from March to August 1996, while 
the upper plots are averages from March 15 to  March 20, 1997.
Both statistical and systematic errors are included in the \hegra plots.  
The solid line corresponds to our model for the spectrum in March 1996, 
the dotted line for the data on  April 7 and the line with small dashes 
for April 16 in 1997.\label{fig.4}}
\end{figure}

\clearpage

\begin{figure}
\epsscale{1.0}
\plotone{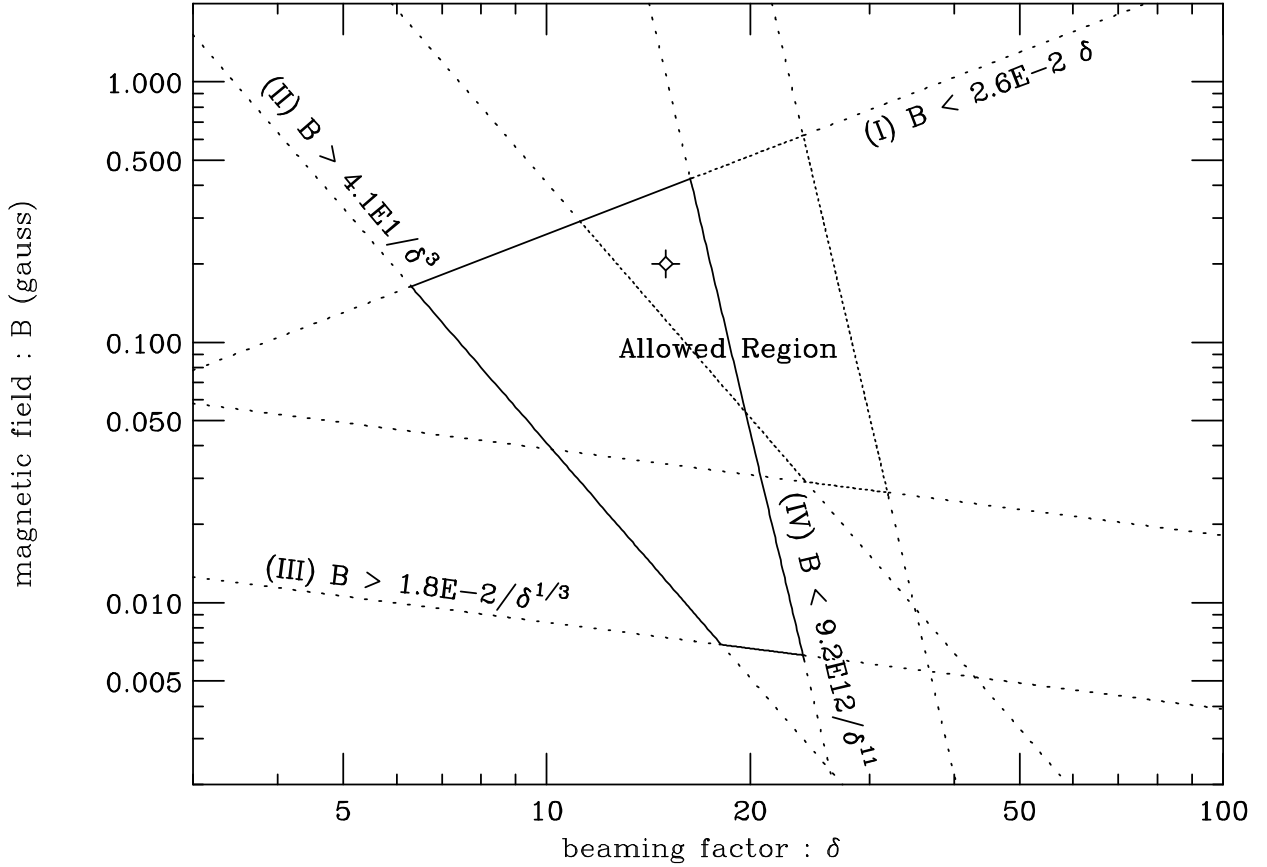}
\figcaption{The parameter space ($B$, $\delta$) allowed by 
the one-zone SSC model for Mrk 501 data. 
The figure surrounded with solid line is the 
allowed parameter space for variability time scale $t_{var}$ = 10$^{5}$ sec 
and the cross point is the parameter used for our model (Figure 4).
The figure with dotted lines is the allowed parameter space 
for $t_{var}$ = 10$^{4}$ sec. Small dashes are constraints from each 
equations described in $\S$4 for the case of $t_{var}$ = 10$^{5}$ sec and 
$t_{var}$ = 10$^{4}$ sec.
(I) derived from the maximum energy of synchrotron emission (assumed here 
     to be 50 keV) and Comptonized spectrum (E $>$ 10 TeV).
(II) derived from the observed luminosity ratio of the synchrotron and 
     Inverse Compton components. We used here $l_{sync}$ $\sim$ $l_{SSC}$   
      $\sim$ 10$^{-10}$ erg cm$^{-2}$ s$^{-1}$.
(III) derived from the evaluation of the synchrotron cooling time ($t_{sync}$) 
      and the variability time scale in the jet frame 
     ($t_{var}$$\delta$).
(IV) derived from the evaluation of the Compton cooling time ($t_{SSC}$) 
     and the variability time scale in the jet frame. 
In the process of (I), we utilize non-contemporaneous data. Note that the 
equation of $\gamma_{max}$ derived in (I) is subsequently used in 
(III) and (IV).\label{fig.5}}
\end{figure}

\end{document}